\documentclass[journal]{IEEEtran}

\makeatletter

\newcommand{\Rmnum}[1]{\expandafter\@slowromancap\romannumeral #1@}
\makeatother
\usepackage{graphicx,cite,epsfig,amssymb,amsmath,amsfonts,multirow}
\usepackage{epstopdf}
\usepackage{mathrsfs}
\usepackage{float}
\usepackage{soul}
\usepackage{colortbl, color}
\usepackage{cases}
\usepackage[ruled,vlined,linesnumbered]{algorithm2e}
\usepackage{verbatim}
\usepackage{url}

\usepackage{booktabs}

\usepackage{ulem}
\newcommand{\ls}[1]  
   {\dimen0=\fontdimen6\the=#1\dimen0
    \advance\lineskip.5\fontdimen5\the\lineskip-\dimen0
    \lineskiplimit=.9\lineskip
    \baselineskip=\lineskip
    \advance\baselineskip\dimen0
    \normallineskip\lineskip
    \normallineskiplimit\lineskiplimit
    \normalbaselineskip\baselineskip
    \ignorespaces
   }

\begin{document}
\bibliographystyle{ieeetr}

\title{Dealing with Limited Backhaul Capacity in Millimeter Wave Systems: A Deep Reinforcement Learning Approach}

\author{Mingjie~Feng,~\IEEEmembership{Student~Member,~IEEE}~and~Shiwen~Mao,~\IEEEmembership{Fellow,~IEEE}%

\thanks{M. Feng and S. Mao are with the Department of Electrical and Computer Engineering, Auburn University, Auburn, AL 36849-5201 USA. Email: mzf0022@auburn.edu, smao@ieee.org.
}
}

\maketitle

\markboth{IEEE Communications Magazine, VOL.XXX, NO.XXX, MONTH YEAR}%
{FENG and MAO: DEALING WITH LIMITED BACKHAUL CAPACITY IN MILLIMETER WAVE SYSTEMS: A DEEP REINFORCEMENT LEARNING APPROACHs}

\pagestyle{headings}\thispagestyle{headings}

\begin{abstract}
Millimeter Wave (MmWave) communication is one of the key technology of fifth generation (5G) wireless systems to achieve the expected 1000x data rate. With large bandwidth at mmWave band, the link capacity between users and base stations (BS) can be much higher compared to sub-6GHz wireless systems. Meanwhile, due to the high cost of infrastructure upgrade, it would be difficult for operators to drastically enhance the capacity of backhaul links between mmWave BSs and the core network. As a result, the data rate provided by backhaul may not be sufficient to support all mmWave links, the backhaul connection becomes the new bottleneck that limits the system performance. On the other hand, as mmWave channels are subject to random blockage, the data rates of mmWave users significantly vary over time. With limited backhaul capacity and highly dynamic data rates of users, how to allocate backhaul resource to each user remains a challenge for mmWave systems. In this article, we present a deep reinforcement learning (DRL) approach to address this challenge. By learning the blockage pattern, the system dynamics can be captured and predicted, resulting in efficient utilization of backhaul resource. We begin with a discussion on DRL and its application in wireless systems. We then investigate the problem backhaul resource allocation and present the DRL based solution. Finally, we discuss open problems for future research and conclude this article.
\end{abstract}

\section{Introduction}

With the explosion of smart devices and data-intensive wireless applications, the demand for high data rate services has drastically increased in recent years. To meet such demand, the fifth generation (5G) cellular network is under intensive research from both industry and academia. According to a recent report, the 5G networks are expected to support massive connections with minimum data rate of 100 Mbps and peak data rate higher than 10 Gbps~\cite{Nokia}. To achieve this goal, several technologies are considered as candidates for 5G systems, including millimeter-wave (mmWave) communication, massive MIMO, and small cell. By operating at mmWave band with large bandwidth, an mmWave system can significantly enhance the data rate performance to multi-Gbps level.

As the data rates of links between an mmWave base station (BS) and users are greatly enhanced, the capacity of backhaul link between the BS and the core network becomes relatively limited, posting a new challenge to mmWave cellular networks. Compared to a long term evolution (LTE) system with typical cell throughput less than 150 Mbps~\cite{Aviat}, the cell through of an mmWave system can be greater than 1.5 Gbps~\cite{Akdeniz14}, which is comparable to the data rate of a current backhaul link. As a result, the backhaul links in mmWave cellular networks are expected to achieve much higher data rates compared to current cellular networks. In current LTE networks, the configuration of a backhaul link is to support peak cell throughput. However, this may not be feasible in mmWave networks. Due to cost concern, it is highly unlikely for operators to upgrade existing infrastructure to drastically enhance capacity of wired backhauls. In case of wireless backahul, e.g., mmWave-based wireless backhaul or free space optical, although the cost can be reduced, the challenge brought by limited backhaul capacity remains. One the one hand, the capacity of wireless backhaul link is shared by multiple BS-user links. On the other hand, the backhaul links are likely experience higher propagation loss than the BS-user links.

The tension caused by limited backhaul capacity may be aggravated in the future as the data rate of mmWave links is expected to keep increasing. For example, high resolution $360^{\circ}$ virtual reality (VR) requires data rate on the order of 1 Gbps and latency of 1 ms. Based on a prediction in~\cite{Nokia}, the 5G mmWave networks need to support 50 Gbps data rate by the 2024. In addition, due to the expected dense deployment of mmWave BSs~\cite{Ge16}, a large number of backhaul connections, which can be wired or wireless, would be coexist. As a result, the achievable data rate of each backhaul link would be limited, which may be caused by resource sharing, mutual interference, potential congestion, or increased overhead~\cite{Feng18}. Therefore, unlike traditional cellular networks (from 1G to 4G) in which the wireless transmission between BS and user is the bottleneck, the backhaul becomes a potential bottleneck in mmWave system. Although some field tests were performed to demonstrate the potential of mmWave cellular systems such as in~\cite{Akdeniz14}, these tests are not based on actually cellular networks. Thus, the impact of limited backhaul capacity has not been tested and verified, which requires further investigation. The challenge of possible bottleneck at backhaul has been observed in the context of ultra-dense small cell deployment~\cite{Ge16}, in which the large number of small cells put pressure on the backhaul links. Compared to the case of network desification, the bottleneck challenge in an mmWave system is caused by the significantly increased data rate of mmWave transmissions.

\begin{figure}[!t]
 \begin{center}
   \includegraphics[width=3.3in]{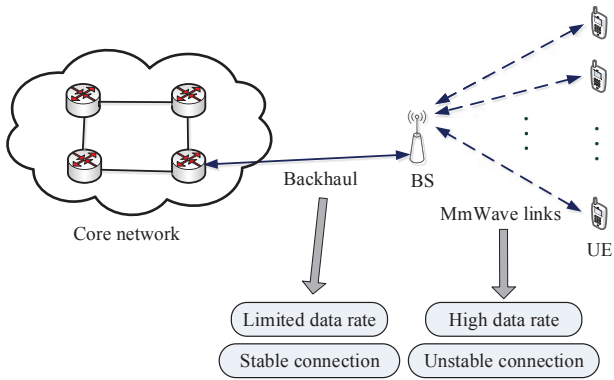}
 \end{center}
\caption{System model of a mmWave system with limited backhaul capacity.}
\label{fig1}
\end{figure}

On the other hand, due to the short wavelength of mmWave communication, the transmissions between BS and users are subject to random blockage. As a result, the data rate of each user is highly dynamic. In contrast, the data rate of a backhaul link is much more stable since it is implement by wired connection or line of sight (LOS) wireless connection. Therefore, the BS-user link is characterized by high data rate and unstable connection, while the backhaul link is characterized by relatively limited data rate and stable connection, as shown in Fig.~\ref{fig1}. To balance such mismatch and enhance the system performance, efficient backhaul resource allocation to each user is necessary. For example, when a user switches from LOS transmission to non line of sight (NLOS) or outage, less resource shoud be allocated to this user. However, such adaptive control cannot be implemented by traditional resource allocation schemes, due to the varying system dynamics. To perform efficient scheduling, a BS needs to predict possible blockage and estimate the data rate of each user based on current channel state information (CSI). Then, it makes decision on the backhaul resource allocation and sends a request to the core network. This way, the backhaul scheduling can be performed in a timely manner that captures the blockage pattern.

Deep reinforcement learning (DRL) is a new paradigm for intelligent decision-making~\cite{Mnih15}, which can be implement by TensorFlow and Keras. Combing reinforcement learning and deep neural network, a DRL agent interacts with the environment and learns the pattern of a Markov Decision Process (MDP) through training experience. Specifically, a DRL agent employs a deep neural network to approximate the Q-values, where the Q-values are defined by discounted cumulative rewards that can be obtained by taking different actions under certain system states. Then, the agent makes optimal decisions based on the estimated Q-values. Compared to other machine learning approaches, DRL is model-free and does not require data samples from an external supervisor. Due to these benefits, the application of DRL is wireless networks has drawn growing attention recently. In this article, we apply DRL to deal with the challenge of limited backhaul capacity in mmWave networks. By learning the blockage pattern based on the CSI of mmWave users, a BS decides the resource allocation of backhaul link with the objective of maximizing the sum utility of all users.

In the remainder of this article, we first introduce the background of DRL and review its recent applications in wireless systems. Then, we present a DRL based approach for backhaul resource allocation. Finally, we discuss open research problems and conclude this article.

\section{Deep Reinforcement Learning for Wireless Systems \label{sec:overview}}

\subsection{Preliminaries of Deep Reinforcement Learning}
A reinforcement learning (RL) agent aims to learn from the environment and take action to maximize the long term cumulative reward. The environment is modeled an MDP with state space $\mathcal{S}$ and a RL agent can take actions from space $\mathcal{A}$. The agent interacts with the environment by taking actions, observing the reward and system state transition, and updating its knowledge about the environment. The objective of a RL algorithm is to find the optimal policy, which determines the strategy of taking actions under certain system states. A policy $\pi$ is specified by $\pi (a|s) = \mathcal{P}\left\{ A_t = a|S_t = s \right\}$. In general, a policy is in a stochastic form to enable exploration over different actions. To find the optimal policy, the key component is to determine the value of each state-action function, also known as Q-function, which is defined by
\begin{align}\label{eq1}
&Q_\pi (s,a) = \mathbb{E}_\pi \left[ G_t|S_t = s,A_t = a \right] \nonumber \\
&\hspace{0.5in}=R_s^a + \gamma \sum\limits_{s'\in S} P_{ss'}^a{v_\pi}(s')
\end{align}
where $R_s^a$ is the instant reward that can be obtained by taking action $a$ under state $s$; $P_{ss'}^a$ is the transition probability from state $s$ to state $s'$ under action $a$; $\gamma$ is the discount factor used to balance the long-term and short-term rewards; $G_t$ is the cumulative reward from time $t$, given by $G_t = \sum_{k = 0}^\infty \gamma ^kR_{t + k + 1} $. In~\eqref{eq1}, $v_\pi (s)$ is the state-value function which indicates the expected reward if the system is in state $s$ and follow policy $\pi$, given by $v_\pi (s) = \mathbb{E}_\pi \left[ G_t|S_t = s \right] = \sum\limits_{a \in A} \pi (a|s) Q_\pi (s,a)$. With Q-functions, an MDP is solved when the optimal policy is found, i.e., $Q_*(s,a) = \mathop {\max }\limits_\pi  Q_\pi (s,a)$. A common RL technique for solving an MDP is Q-learning, which uses an empirical iterative approach to update the values of Q-functions (Q-values). In particular, an agent interacts with the environment by taking actions and obtaining reward, and then update the Q-values by
\begin{align}\label{eq2}
&Q(s_t,a_t) \leftarrow Q(s_t,a_t) \nonumber \\
&\hspace{0.5in}+ \alpha \left[ R_{t + 1} + \gamma \mathop {\max }\limits_{a_{t + 1}} Q(s_{t + 1},a_{t + 1}) - Q(s_t,a_t) \right]
\end{align}

RL has been applied in decision-making problems of mmWave networks such as in~\cite{Mezzavilla16}. However, in large scale systems with large numbers of states and actions, traditional Q-learning approach becomes infeasible since a table is required to store all the Q-values. In addition, traditional Q-learning needs to visit and evaluate every state-action pair, resulting in huge complexity and slow convergence. An effective approach to deal with such challenge is use a neural network (NN) to approximate the Q-values, given by $Q(s,a,{\bf{w}}) \approx {Q_\pi }(s,a)$, where $\bf{w}$ are the weights of the NN. By training a NN with sampled data, the NN can map the inputs of state-action pairs to their corresponding Q-values. However, a direct application of NN in Q-learning may be unstable or even diverge due to the correlations between training samples and the correlations between Q-values and target values~\cite{Mnih15}.

\begin{figure}[!t]
 \begin{center}
   \includegraphics[width=3.3in]{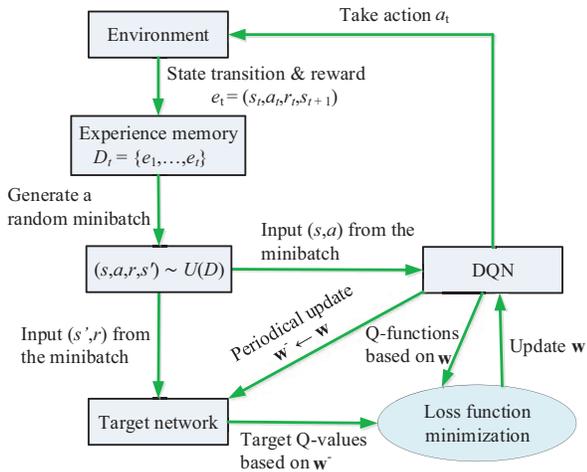}
 \end{center}
\caption{Framework of the DRL approach in~\cite{Mnih15}}
\label{fig2}
\end{figure}

To reduce such correlations, a DRL approach was proposed in~\cite{Mnih15}, in which a deep neural network (DNN) is used to approximate the Q-value, yielding a deep Q-network (DQN). In the DRL approach presented in~\cite{Mnih15}, the agent first explores the environment by randomly taking actions and stores the experience, $e_t = (s_t,a_t,R_t,s_{t + 1})$, in a target network. Then, a mechanism called {\em experience replay} is used, where the data are randomly sampled in minibatches from the target network to break the correlation in a sequence of observation. With the samples from the target network, the weights of the DQN is updated by minimizing the loss function given by
\begin{align}\label{eq3}
L_i(w_i) = \mathbb{E}_{(s,a,r,s') \in U(D)}\left[ \left( {r + \gamma \mathop {\max }\limits_{a'} Q(s',a',w_i^ - )} \right. \right. \nonumber \\
\left. {\left. { - Q(s,a,{w_i})} \right)}^2 \right]
\end{align}
where $w_i$ and $w_i^-$ are the weights of DQN and target network at iteration $i$, respectively. The loss function~\eqref{eq3} is the mean square error between DQN and target network, which can be minimized through stochastic gradient descent. To reduce the correlation between DQN and target network, the target network is updated less frequently. After the training of DQN, the agent then takes action based on the estimated Q-values. The general framework of the DRL approach in~\cite{Mnih15} is shown in Fig.~\ref{fig2}.

\subsection{Applications in Wireless Networks}

In the design of wireless networks, a major challenge is to solve the formulated combinatorial problems. While exhaustive search is infeasible due to prohibitive complexity, existing solutions typically rely on network information exchange, which yields a tradeoff between overhead and performance. For DRL approaches, the network optimization is based on trial and error processes, which does not require explicit or instantaneous network information. In particular, a DRL algorithm is model-free, which does not require explicit knowledge on the inter-dependent patterns of different nodes. In addition, with extensive offline training, a DRL agent is able to predict the system dynamics, which enables timely scheduling. Thus, compared to traditional approaches, DRL-base schemes have the potential to achieve better performance with reduced online overhead.

\begin{table*}
\vspace{0.5in} \centering \caption{Applications of DRL in Different Wireless Networks} \label{tab1}
\begin{tabular}{c|c|c|c|c|c}
\toprule
\multirow{2}*{} & \multirow{2}*{\bf{Application}} & \multirow{2}*{\bf{State}} & \multirow{2}*{\bf{Action}} & \multirow{2}*{\bf{Reward}} & \multirow{2}*{\bf{Learning Objective}}  \\ & & & & & \\
\midrule
\multirow{2}*{\cite{He17}} & Cache based & \multirow{2}*{Channel power gain} & User selection for & \multirow{2}*{Network throughput} & Channel dynamics\& \\ & interference alignment &  & interference alignment & & cache availability \\
\midrule
\multirow{2}*{\cite{Wang18}} & Multi-channel & Channel state:  & Channel selection & Number of & \multirow{2}*{Channel availability} \\ & access & good/bad & of each user & successful transmissions & \\
\midrule
\multirow{2}*{\cite{Challita18}} & Resource management & Current channel & Channel access & Total throughput & Channel access patterns \\ & in LTE-Unlicensed & usage pattern & probability & on selected channels & of other users \\
\midrule
\multirow{2}*{\cite{Wang2018}} & Handover control in & Signal qualities from & \multirow{2}*{BS selection} & Weighed sum of data & Prediction for channel qualities \\
& ultra-dense network & different BSs &  &  rate \& handover energy &  from different BSs\\
\midrule
\multirow{2}*{\cite{Xu18}} & Traffic allocation & Throughput \& delay & \multirow{2}*{Traffic split ratio} & Total utility (weighted & Learn traffic pattern\\ & in multi-hop network & of each session & & sum of throughput \& delay) & from experience \\
\midrule
\multirow{2}*{\cite{Naparstek17}} & Multi-channel & Channel access & \multirow{2}*{Channel access strategy} & Number of & Probabilities of success\\ & random access & of other users &  & successful transmissions & transmission over multi-channel \\
\bottomrule
\end{tabular}
\end{table*}

Due to such promising prospect, DRL algorithms have been recently considered in several wireless networks to perform intelligent decision making~\cite{He17,Wang18,Challita18,Wang2018,Xu18,Naparstek17}. In~\cite{He17}, DRL is used to estimate the availability of cache and select proper set of users for interference alignment. In~\cite{Wang18,Naparstek17,Challita18}, the problem of multi-channel access is considered in which each user observes the channel dynamics from history and estimates the possible actions of other users, then determines its channel access strategy. In~\cite{Wang2018}, DRL is used to predict the QoS that can obtained when handover to another BS, resulting in efficient handover process. In~\cite{Xu18}, continuous actions and states are considered so that DQN-based DRL cannot be applied. The deep deterministic policy gradient (DDPG), which is based on actor-critic framework, was employed to address continuous space control problem. The general idea is to parameterize the Q-functions and derive the optimal values of parameters through policy gradient. In~\cite{Challita18,Wang2018,Naparstek17}, the problems are formulated as multi-agent control with interactions among agents. As a result, experience replay for a single agent cannot be applied in such scenarios. To take the inter-agent impact into account, the long short term memory (LSTM) approach is used to generate target values. The key aspects of system models in recent works are summarized in Table~\ref{tab1}.

\section{DRL Based Backhaul Resource Allocation \label{sec:approach}}
\subsection{System Model}
We consider an mmWave BS serving $K$ user equipments (UE) indexed by $k=1,...,K$. Each UE has three link states, LOS, NLOS, and outage, which are denoted by three binary 0-1 variables, $x_{k,i}(t)$, $i=1,2,3$. Specifically, $x_{k,1}(t)=1$, $x_{k,2}(t)=1$, and $x_{k,3}(t)=1$ indicate that user $k$ is under LOS, NLOS, and outage state at time $t$, respectively. The link state of each user follows a Markov process with steady probabilities given in~\cite{Akdeniz14}. We assume that the BS can estimate the values of $x_{k,i}(t)$ through the statistics of user signals. The BS can also measure the {\em achievable} data rate of mmWave link for user $k$, $C_k(t)$ , via uplink signal.

We assume the backhaul resource is divided into $M$ orthogonal blocks, $M>K$, each block can be a period of time or a range of wavelength. The capacity of each block is $U$, then the total backhaul capacity is $U\cdot M$. Let $n_k(t)$ be the number of blocks allocated to user $k$, the backhaul capacity allocated to user $k$ is $B_k(t)=U\cdot n_k(t)$. Then, the {\em actual} data rate of user $k$ is $R_k(t) = \min \left\{ B_k(t),C_k(t) \right\}$.

\subsection{DRL Framework}
\begin{figure}[!t]
 \begin{center}
   \includegraphics[width=3.4in]{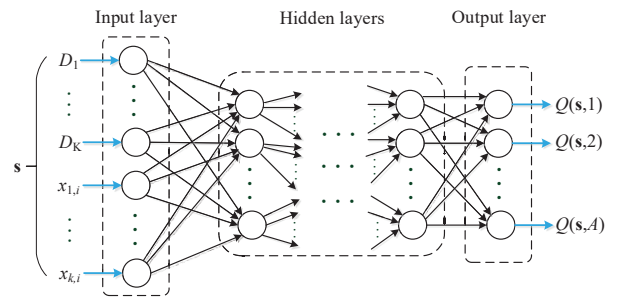}
 \end{center}
\caption{Architecture of the DQN for backhaul resource allocation.}
\label{fig3}
\end{figure}

The proposed DRL-based approach employs a DQN to find the resource allocation strategy under different system states. The key component of system state is the achievable data rate of each UE, $C_k(t)$. We also set the link state of each UE, $x_{k,i}(t)$, as part of the system states, since it impacts the future data rates. Then, the system state is used as input of the DQN. The action taken by the agent indicates the backhaul capacity allocation, i.e., the number of blocks allocated to each user $n_k(t)$. The action space is consisted of all feasible resource allocation, which includes multiple combinations of integers that satisfy $\sum_k n_k(t) = M$, and we index the actions by $a(t)=1,...,A$. To achieve a good system performance as well as guarantee the fairness among users, we define the utility of each user to be a concave function of its data rate. Then, the system reward is set as the sum of utilities of all users. The architecture of the DQN is shown in Fig.~\ref{fig3}. The input layer includes the link state and achievable data rate information of all UEs. The output layer presents the approximated Q-values and there are several hidden layers between input and output layers. To match the capacity of a backhaul resource block, we define $D_k(t) = \left\lceil \frac{C_k(t)}{U} \right\rceil$ and use it at the input layer of the DQN. $D_k(t)$ indicates the numbers of resource blocks needed to satisfy the data rate requirements of UE $k$.

The training procedure of the DQN is the same as the one in~\cite{Mnih15}, which uses experience replay to reduce the correlation between training samples, as shown in Fig.~\ref{fig2}. With the DQN, the agent at the BS first observes the current system state, i.e., the values of $D_k(t)$ and $x_{k,i}(t)$ of all users. Then, it obtains the Q-values of taking different actions, i.e., selecting different resource allocation strategies. With the Q-values, the agent takes an action according to the $\epsilon$-greedy approach, which selects the action with maximum Q-value with probability $1-\epsilon$ and randomly selects an action with probability $\epsilon$.

\subsection{Illustrative Example}
We evaluate the performance of the DRL based approach with simulations. We consider an mmWave cell with a coverage radius of 100 m, users are randomly distributed in the cell. Let $d$ be the distance between a user and the BS,
The probabilities of a user in different link states are functions of $d$. The probabilities under outage, LOS, and NLOS are $p_{\rm{out}}(d)=\max(0,1-e^{-a_{\rm{out}}+b_{\rm{out}}})$, $p_{\rm{LOS}}(d)=(1-p_{\rm{out}}(d))e^{-a_{\rm{los}}d}$, and $p_{\rm{NLOS}}(d)=1-p_{\rm{out}}(d)-p_{\rm{LOS}}(d)$, respectively~\cite{Akdeniz14}, which are the steady probabilities of the Markov process of link state $x_{k,i}(t)$. We employ the channel model of 73 GHZ band in~\cite{Akdeniz14}, where the NLOS links experience higher path loss than the LOS links. The system bandwidth is 1 GHz, the transmission powers of BS and UEs are 30 dBm and 20 dBm, respectively. The backhaul capacity is 10 Gbps, the backhaul resource is divided into 20 resource blocks. There are two hidden layers in the DQN and we use ReLu as the activation function. We consider two DRL-based schemes, namely DRL-1 and DRL-2, with reward functions given as $\sum_k \log (R_k(t))$ and $\sum_k \sqrt {R_k(t)}$, respectively. With logarithmic utility function, DRL-1 scheme achieves proportion fairness. Compared to DRL-1, DRL-2 is more efficiency-prone with worse fairness. Two benchmark schemes are considered for comparison, a myopic scheme and the equal allocation scheme. For the myopic scheme, the backhaul resource allocation is based on current data rates of mmWave links, without considering the future change of link states.

\begin{figure}[!t]
 \begin{center}
   \includegraphics[width=3.3in]{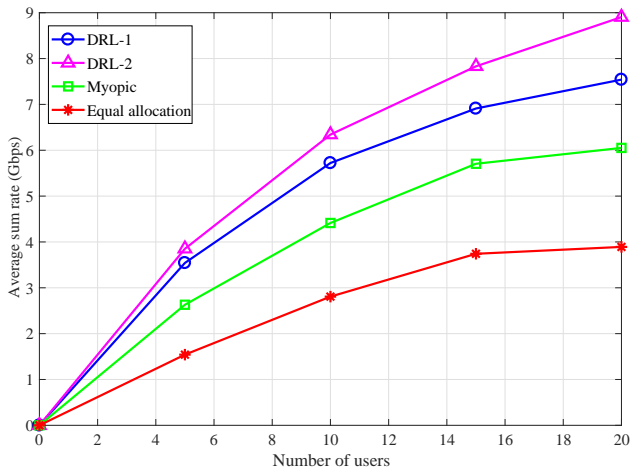}
 \end{center}
\caption{Sun rate performance of different schemes versus the number of users.}
\label{fig4}
\end{figure}

Fig.~\ref{fig4} shows the sum rate performance under different number of users. As the number of users increases, the sum rates of all schemes grow at reduced rates, showing that the system performance is limited by the backhaul capacity. The proposed DRL-based schemes outperforms other ones and the performance gap is enlarged when the number of users increases. This is because the BS is able to predict the variation of link state and allocate the resource based on long-term consideration. Then, the backhaul resource can be efficiently utilized, and such advantage becomes significant when the number of users is large. Compared to DRL-1 scheme, DRL-2 achieves higher data rate since its utility and reward functions are set to prioritize efficiency over fairness.

\begin{figure}[!t]
 \begin{center}
   \includegraphics[width=3.3in]{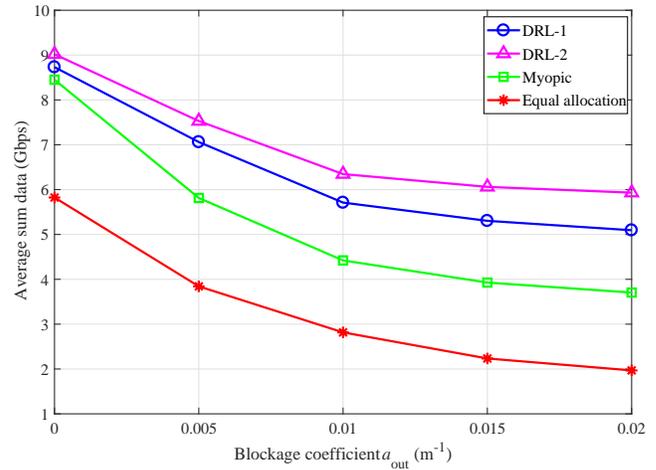}
 \end{center}
\caption{Performance of different schemes versus blockage coefficient $a_{\rm{out}}$.}
\label{fig5}
\end{figure}

The performance under different values of blockage coefficient $a_{\rm{out}}$ is shown in Fig.~\ref{fig5}. The blockage coefficient is defined in~\cite{Akdeniz14}, which indicates the likelihood that a user experience blockage. Given the same BS-UE distance, a scenario with larger $a_{\rm{out}}$ has higher blockage probability compared to a scenario with lower $a_{\rm{out}}$. From Fig.~\ref{fig5}, we can see that when $a_{\rm{out}}$ is small, the performance of the myopic scheme is close to the proposed DRL-based schemes, since the ratio of of users under blockage is small and the data rates of mmWave links are relatively stable. However, when the number of users increases, the performance gap between the proposed schemes and the myopic scheme is increased, showing that DRL-based schedule is effective in capturing the system dynamics and making intelligent decisions from the perspective of long-term benefit.

\section{Open Problems and Future Research \label{sec:future}}

\subsection{Joint Optimization of Backhaul and MmWave Link}
The DRL based backhaul resource allocation presented in Section~\ref{sec:approach} is based on given achievable rate of each user. To mitigate the pressure caused by limited backhaul capacity, the design of BS-user links can also be considered. The design of resource allocation in LTE systems with limited backhaul capacity has been studied in~\cite{Hg12}. In mmWave systems, the data rate of each mmWave link can be adjusted through precoding design. Considering the channel characteristics of different users, a joint consideration of backhaul resource allocation and precoding can provide a better solution to balance the tension between limited backhaul and increased mmWave data rate demand.

\subsection{Dynamic Backhaul Capacity}
In our model, we assume fixed capacity for backhaul, which corresponds to the case of wired backhaul or LOS mmWave backhaul with highly stable data rate. However, in a practical system with wireless backhaul, the data rate of backhaul would vary over time. Thus, it is necessary for the agent to learn such dynamics as well, and more sophisticated design is required based on the proposed framework.

\subsection{Multi-Cell Scenario}

\subsubsection{Capacity Allocation Among Different Backhauls}
The design in Section~\ref{sec:approach} is based on a single cell scenario. From the perspective of multi-cell, the capacity allocated to each backhaul can be optimized to further enhance the system performance. For example, an mmWave BS with heavy traffic and high aggregated data rate requirement can share more capacity from the core network. However, the load balancing and capacity allocation require coordination between different BSs and efficient design is required. In addition, how to address the scalability issue would be another challenge. Capacity allocation among different backhauls for load balancing has been investigated in other wireless networks, such as in heterogeneous cloud radio access networks~\cite{Ran15}. Due to the dynamic nature of mmWave communications, the varying capacity requirement of each backhaul need to be learned to enable effective scheduling.

\subsubsection{Adaptive User Association}
To mitigate the pressure of limited backhaul, an effective approach is to perform load balancing. For a BS with large deficit in backhaul capacity, part of the users severed by the BS can handover to neighboring BSs to reduce the traffic demand on this BS. Thus, traffic-aware user association is another design factor that can considered for better system performance.

\subsection{Heterogeneous Network}
In a heterogeneous network, the traffic of small cells is transmitted to a macrocell via backhaul connections and then forwarded to the core network via the backhaul of the macrocell. Then, the backhaul resource allocation becomes a two-tier problem, which requires more complicated design. In addition, similar to the multi-cell case, the capacity allocation for different small cell backhaul links and adaptive user association are important design issues that should be jointly considered with backhaul resource allocation.

\subsection{Caching Assisted System}
BS caching, e.g., femtocaching, was recently proposed as an effective approach to enhance the data rate of users. By downloading popular contents in advance and storing at local BSs, the files requested by users are directly transmitted from local BS. While the primary goal of caching is to increase the capacity of BS-user links and reduce delay, it is also a good solution to the limited backhaul capacity challenge. When the traffic load of an mmWave BS is low, it can request popular files from the core network. When the traffic load is increased, the popular files at the BS can be used to satisfy the demand of some users. As a result, the backhaul capacitiy is mainly used to satisfy the instantaneous demands from users, thus mitigating the traffic burden at the backhaul. Under the caching architecture, the key design issue is the selection of popular contents. With limited storage, it is necessary to learn the patterns of users preference and blockage. For example, when a users is under frequent blockage, caching and storing the content of this user would lead to under-utilization. However, if the content requested by the user is also frequently requested by other users, the utilization of would be improved. Thus, the agent needs to learn multiple patterns to derive an efficient caching strategy.

\subsection{Performance-Complexity Tradeoff}
In the system model of Section~\ref{sec:approach}, we assume the backhaul resource is divided into $M$ blocks. To improve resource utilization and enhance the system performance, a larger value of $M$ is desirable. However, this results in increased dimensions of both action and state spaces. Thus, an adaptive selection of $M$ that achieves a good tradeoff between complexity and performance is another design issue.

\section{Conclusion}
In this article, we aim to address the challenge of limited backhaul capacity in mmWave networks with a DRL based approach. We first overview the background of DRL and its applications in wireless networks. Then, we present a DRL based approach to enable efficient backhaul resource allocation, and show the effectiveness through an illustrative example. We then discuss the future research problems and conclude this article.

\section*{Acknowledgment}

This work was supported in part by the NSF under Grant CNS-1702957 and by the Wireless Engineering Research and Engineering Center at Auburn University.


\bigskip
\noindent
{\bf Mingjie Feng} [S'15] received his Ph.D. degree in Electrical and Computer Engineering from Auburn University, Auburn, AL, USA, in 2018. He received his Bachelor's and Master's degrees from School of Electronic Information and Communications, Huazhong University of Science and Technology, Wuhan, China, in 2010 and 2013, respectively. He is currently a postdoctoral research associate in the Department of Electrical and Computer Engineering at the University of Arizona. In 2013, he was a visiting student in the Department of Computer Science and Engineering, Hong Kong University of Science and Technology. His research interests include mmWave communication, massive MIMO, cognitive radio networks, heterogeneous networks, and full-duplex communication. He is a recipient of a Woltosz Fellowship at Auburn University.

\vfil

\noindent
{\bf Shiwen Mao} [S'99-M'04-SM'09-F'19] received his Ph.D. in ECE from Polytechnic University, Brooklyn, NY in 2004. He is the Samuel Ginn Distinguished Professor and Director of the Wireless Engineering Research and Education Center at Auburn University, Auburn, AL. His research interests include wireless networks and multimedia communications. He is a Distinguished Speaker of IEEE Vehicular Technology Society. He received the 2017 IEEE ComSoc ITC Outstanding Service Award, the 2015 IEEE ComSoc TC-CSR Distinguished Service Award, the 2013 IEEE ComSoc MMTC Outstanding Leadership Award, and the NSF CAREER Award in 2010. He is a co-recipient of IEEE ComSoc MMTC Best Conference Paper Award in 2018, Best Demo Award from IEEE SECON 2017, Best Paper Awards from IEEE GLOBECOM 2016 \& 2015, IEEE WCNC 2015, and IEEE ICC 2013, and 2004 IEEE Communications Society Leonard G. Abraham Prize in the Field of Communications Systems. He is an IEEE Fellow.s

\end{document}